\documentstyle[seceq,epsf,amssymb]{ptptex}

\def\bea{\begin{eqnarray}}
\def\ena{\end{eqnarray}}
\renewcommand{\a}{\alpha}

\renewcommand{\c}{\gamma}
\renewcommand{\d}{\delta}
\newcommand{\s}{\sigma}

\newcommand{\e}{\epsilon}

\newcommand{\pa}{\partial}
\newcommand{\nn}{\nonumber\\}
\newcommand{\lan}{{\langle}}
\newcommand{\ran}{\rangle}
\newcommand{\vs}[1]{\vspace{#1 mm}}
\newcommand{\hs}[1]{\hspace{#1 mm}}

\newcommand{\p}[1]{(\ref{#1})}

\title{
Introduction to branes and M-theory for relativists and cosmologists\footnote{
Lectures at the international workshop ``Brane world''
at YITP, 15--18 January 2002.~\cite{NO1}}}

\author{%       %Use \sc for the family name
Nobuyoshi {\sc Ohta}\footnote{e-mail address: ohta@phys.sci.osaka-u.ac.jp}
}

\inst{
Department of Physics, Osaka University,
Toyonaka 560-0043, Japan
}

\preprintnumber{OU-HET 411 \\ gr-qc/0205036}

\recdate{%      %Editorial Office will fill in this.
\today
}

\abst{
We review the recent developments in superstrings. We start with a brief
summary of various consistent superstring theories and discuss T-duality
which necessarily leads to the presence of D-branes. The properties of
D-branes are summarized and we discuss how these suggest the
existence of 11-dimensional quantum theory, M-theory, which is believed
to give rise to various superstrings as perturbative expansions around
particular backgrounds in the theory. We also discuss the interpretation of
brane solutions as black holes in string theories and statistical explanation
of Bekenstein-Hawking entropy. The idea behind this interpretation is that
there is a fundamental duality between closed (gravity) and open (gauge theory)
string degrees of freedom, one of whose manifestation is what is kown as
AdS/CFT correspondence. The idea is used to discuss the greybody factors
for BTZ black holes. Finally the entropy of various balck holes are
discussed in connection with Cardy-Verlinde formula.
}

\begin{document}

\maketitle

\tableofcontents

\section{Introduction}

One of the long standing problems in particle physics and gravitational
theories is how to understand quantum theory of gravity. It is notoriously
difficult to make sense of quantum theory of gravity. The only possible
candidate for this is the superstring theory which seems to exhibit good
perturbative behavior. However, it has been known that there are, at least,
five distinct consistent superstring theories, and it seems that they
simply exist without any relation between them. If they exist as such, it
is extremely difficult to determine which particular theory describes our
real world. The recent developments in the nonperturbative understanding of
string theories begin casting light to this important question.

The new developments started with the discovery of various extended objects
in superstring theories, among which the most important are the so-called
Dirichlet-branes (D-branes for short).~\cite{PO1} It has become clear that
they play very important roles in understanding strong coupling dynamics of
superstrings, and in particular lead to M-theory notion.~\cite{WI1}
Superstring theories, when viewed in the strong coupling, are not just
theories of strings but they contain many extended objects (branes) as
light degrees of freedom, and the very existence of these objects turned out
to be the origin of the dual relations of apparently different superstring
theories. M-theory, as it is called now, is an 11-dimensional quantum theory
of vastly many extended objects which produces all superstring theories
around its perturbative vacua. Our first aim is to explain how this picture
comes about in the recent developments.

D-branes play very important roles not only in the above picture of M-theory
but also many other places. There are two rather seemingly different
descriptions of D-branes; one is as classical solutions in the low-energy
effective theories, supergravities,\footnote{For brane solutions in the
low-energy effective supergravity, see refs.~\citen{DKL}.}
and the other is in terms of perturbative open strings. Note that these
methods are quite different because the former is in terms of closed string
(gravity) degrees of freedom and the latter is in terms of open string
(gauge theory). The difference is understood as that in the strength of
string coupling constant. The second description is of course valid in the
weak string coupling. The first description allows the interpretation of
D-branes as black hole solutions when suitably compactified.~\cite{STE}
The BPS properties of D-branes (which means that they preserve partial
supersymmetry and hence they are protected from quantum corrections) is
then used to argue that the degrees of freedom in the solutions are the
same in both pictures. It was found that this gives the remarkable
explanation of the black hole entropy in the statistical
mechanics.~\cite{SV,CM}

The success of the description of black holes in terms of open string degrees
of freedom suggests fundamental duality between open and closed strings,
which is deeply tied with the early suggestion of 't Hooft on the connection
between gauge and string theories. ``Duality'' is a word which means that
there are two apparently different descriptions of the same system.
Additional proposal has also been made on the duality between open and
closed strings by Maldacena under the name of AdS/CFT
correspondence.~\cite{MAL,WI3,GKP}\footnote{For a review with extensive
references, see ref.~\citen{AGM}.} This kind of duality appears in many
places in string theories. It has even been suggested that this AdS/CFT
correspondence might be useful in nonperturbative formulation of string
or M-theory.

A related suggestion is the proposal of the matrix model of the
M-theory,~\cite{BFSS} which is a model to describe the theory by $U(N)$
super Yang-Mills (SYM) theory in the infinite momentum frame.
Though there are accumulating evidences of the existence of M-theory,
there has been no convincing proposal on how to formulate the M-theory
itself, and the matrix model is the only candidate we have at the moment.

Motivated by these developments in string theories, brane world scenario has
been suggested and actively studied.~\cite{RS} Here branes are regarded
as the world we are living in, and it is hoped that knowledge of branes,
especially orientifolds, are useful for further elaboration in this area
of research.

This review tries to clarify the question what is D-branes and M-theory
for those who are not familiar with superstrings. The subjects we will review
are:
\begin{itemize}
\item
What is perturbative string theory.
\item
What is the strong coupling limit of type II theory --- M-theory.
\item
How all the string theories are related in this M-theory context.
\item
Duality between open and closed string degrees
of freedom --- the so-called AdS/CFT correspondence.
\item
Applications of AdS/CFT correspondence and holographic principle
\end{itemize}

In \S~2 to 6, string theories and the T-duality are quickly reviewed.
In \S~7, the first appearance of M-theory as the strong coupling limit of
type IIA superstring theory is discussed, together with its relation to
type IIB superstring. The following \S~8 to 10 discuss the whole picture
of the relation of superstring theories, the so-called string web.
In \S~11, we discuss the black hole entropy problem and in \S~12 AdS/CFT
correspondence. The rest of the paper is devoted to the applications of these
general ideas to the computation of greybody factors, description of
noncommutative theories and Cardy-Verlinde formula for black hole entropy.
Conclusions are given in \S~16.

More detailed account of these subjects can be found in the
reviews~\citen{NO2,PO2,TOW,NO3}.

\section{Quick Review of Closed String Quantization}

Let us start with the worldsheet action
\bea
S=-\frac{1}{2\pi\a'}\frac12\int dt d\sigma \sqrt{-g}g^{ij}
\pa_i X^\mu \pa_j X^\nu \eta_{\mu\nu},
\label{action}
\ena
where $1/2\pi\a'$ is the string tension, $i,j (=0,1)$ denote the worldsheet
coordinates $t$ and $\s$, and $X^\mu$ are the space-time coordinates with
$\mu$ and $\nu$ running over $0,\ldots,D-1$. This system can be understood
as 2-dimensional gravity coupled to ``matter'' $X^\mu$.

Properties of this action are:
\begin{itemize}
\item It possesses 2-dimensional reparametrization invariance.
\item It is invariant under the Weyl transformation $g_{ij} \to f g_{ij}$.
\end{itemize}
These invariances allow us to choose $g_{ij}=\eta_{ij}$.

For closed strings obeying periodic boundary conditions, the field equations
following from the action~\p{action} give the {\it mode expansions} for
coordinates:
\bea
X^\mu &=& x^\mu + \a' p^\mu t + i\sqrt{\frac{\a'}{2}}\sum_{n\neq 0}
\Big( \frac{\a_n^\mu}{n} e^{-in(t-\s)} + \frac{\tilde \a_n^\mu}{n}
 e^{-in(t+\s)} \Big) \nn
&=& x^\mu + \frac{\a'}{2} p^\mu \ln|z|^2 + i\sqrt{\frac{\a'}{2}}\sum_{n\neq 0}
\Big( \frac{\a_n^\mu}{n} z^{-n} + \frac{\tilde \a_n^\mu}{n} \bar z^{-n} \Big).
\ena
where $z \equiv e^{i(t-\s)}=e^{\tau-i\s}$. Those with and without tildes are
called {\it right-} and {\it left-movers}, respectively.

For superstrings we also need {\it 2-dimensional fermions}:
\bea
\psi^\mu = \left(\begin{array}{c}
\psi^\mu \\
\tilde\psi^\mu
\end{array}\right),
\ena
which are real (Majorana) and satisfy either periodic or anti-periodic boundary
conditions:
\bea
\psi^\mu(\s+2\pi) = \pm \psi(\s)^\mu: \begin{array}{l}
{\rm R(amond)} \\
{\rm N(eveu)-S(chwarz)}
\end{array},
\ena
and similarly for right-movers.
They are called {\it R (Ramond) or NS (Neveu-Schwarz) sectors}, as indicated.
The {\it mode expansions for fermions} are given as
\bea
\psi^\mu = \left\{\begin{array}{lcc}
\displaystyle{\sum_{m \in \mathbb{Z}}} d_m^\mu z^{-m} &:& {\rm R} \\
\displaystyle{\sum_{r \in {\mathbb Z}+1/2}} b_r^\mu z^{-r} &:& {\rm NS}
\end{array}\right.
\ena

The usual procedure of {\it quantization} leads to the commutation relations
between these modes:
\bea
&& [\a_m^\mu, \a_n^\nu] = m\eta^{\mu\nu}\d_{m+n,0}, \nn
&& \{b_r^\mu, b_s^\nu\} = \eta^{\mu\nu}_{r+s,0}, \quad
\{d_m^\mu, d_n^\nu\} = \eta^{\mu\nu}_{m+n,0},
\ena

For the no-ghost theorem to be valid, the space-time dimension
must be 10 for this fermionic string. This is known as the
{\it critical dimension}.~\cite{NO5}

The ground states for each sector are as follows:
\vs{1}
\begin{description}
\item[NS:]\ \
The ground state is defined by $b_r^\mu |k\ran =0$ for $r\geq \frac{1}{2}$.\\
This means that it is a scalar state. Actually it gives a tachyon, which
should be projected out and we are left with the next massless tensor states.
\vs{-2}
\item[R:] \ \
It is defined by $d_m^\mu |k \ran =0$ for $m \geq 1$. \\
Here we also have the zero modes $d_0^\mu$ satisfying $\{d_0^\mu, d_0^\nu \}
=\eta^{\mu\nu}$, which means that they are actually 10-dimensional $\c$
matrices. The energy of the ground state does not change under the action of
these zero modes. Therefore the ground state must be a representation of
the $\c$ matrices, which is a 10-dimensional spinor. Thus we find that
this sector gives a space-time fermion. Note that the irreducible
10-dimensional spinors are Majorana-Weyl.
\end{description}

These ground states exist for both left- and right-movers, and we have
towers of massive states constructed by multiplying non-zero modes on
the ground states. To make closed string we have to multiply both (left-
and right-)movers. Each mover has supersymmetry. So there are two ways to
make the closed superstring depending on the chiralities of the fermions of
each mover.

If we multiply opposite chiralities $\psi_+, \tilde\psi_-$, we get
a vector-like theory known as {\it type IIA superstring}.

If we multiply same chiralities $\psi_+, \tilde\psi_+$, we get a chiral
theory known as {\it type IIB superstring}.

In addition, states in the theories are chosen by GSO projection
$G\equiv \frac{1+(-1)^F}{2}$, which in fact projects out the tachyon in
the NS sector and keeps only Majorana-Weyl fermions in the R sector,
matching the degrees of freedom in both sectors.

Other states in the theories are as follows:\\
(NS, NS): yields massless tensor fields of 2nd-rank (graviton, dilaton, 2-form)
and massive states on them. \\
$\left. \hs{-2} {\begin{array}{c}
{\rm (NS, R)} \\
{\rm (R, NS)}
\end{array}}
\right\}$ yields fermion fields, namely two chiral gravitini with chiralities
same as the supersymmetry, and massive states on them. This gives
vector-like type IIA theory with gravitini of opposite chiralities and
chiral type IIB theory with those of same chiralities. \\
(R, R): yields massless antisymmetric tensors $F_n \equiv \tilde \psi_\pm^T
 \c_{\mu_1 \cdots \mu_n} \psi_+$.\footnote{It should be noted that these are
field strengths but not potentials.} Due to the kinematics, some of these
vanish, leaving $n=2,4$ for type IIA theory and $n=1,3,5$ for type IIB theory.
These theories possess $N=2$ supersymmetry coming from left- and right-movers,
and hence the name type II.

{\it Heterotic string theories} are constructed by using the above superstring
for, say, the left-mover and bosonic string for right-mover. It turns out
that the consistency of the resulting theories requires that the theories
be restricted only to those with $E_8\times E_8$ or $SO(32)$ gauge symmetries.
Since only the left-mover has supersymmetry, these theories have only
$N=1$ supersymmetry in 10 dimensions.

All of these are theories of closed strings only.
Other consistent theory is $SO(32)$ type I theory which contains open strings
as well. We now discuss this fifth and final consistent superstring theory
briefly.

\section{Open String}

When we consider the variation of the action~\p{action} for open strings,
there arises surface terms from the boundary. We must require that it vanish:
\bea
\d X^\mu \pa_\s X_\mu=0,
\label{obc}
\ena
which means either $\pa_\s X^\mu =0$ (Neumann) or $\d X^\mu=0$ (Dirichlet).

Usually Neumann (free) boundary condition is employed, and this yields open
string. The mode expansion is simply the same as the closed string with
the constraint that the left- and right-movers are the same.
The ground state is spin 1 gauge (vector) particle plus gaugino.
It is possible to attach internal degrees of freedom at the end points
of the open strings, known as {\it Chan-Paton factors}. Consistency of the
superstring requires that the resulting gauge group should be $SO(32)$.
The nature of this gauge group indicates that the strings have no
orientation. Also due to the boundary condition~\p{obc}, left- and right-movers
are not independent and only $N=1$ supersymmetry remains in this
theory. Since the quantum theory of strings necessarily includes closed
strings, this gives a coupled theory of unoriented open and closed strings
with $N=1$ supersymmetry, known as {\it $SO(32)$ type I superstring}.

On the other hand, we could also consider Dirichlet (fixed) boundary condition.
This is a new condition which was not considered before because the boundary
condition violates translational invariance, and hence momentum conservation.
This means that there must be something at the boundary which makes
the momentum conserved. These objects are now called D-branes
because they are defined by the Dirichlet condition of the
end points of the open string.

Conversely D-branes are defined by the property that open strings
can attach to. Obviously these objects break half of the supersymmetry,
just as the open strings, and gives BPS states.

Here we have discussed only the boundary conditions for bosonic part $X^\mu$,
but of course it is possible to impose similar boundary conditions on
fermions. These give consistent definitions of D-branes.

It should be emphasized that the D-branes are not just something we can
consider, but these are objects that we {\it must} consider because,
as we will see shortly, T-duality in string theory forces us to include them.
Moreover, it turns out that they have very important properties to be
summarized later.

\section{T-duality of Closed String}

Consider closed string theory compactified on a circle of radius $R$:
\bea
X \simeq X + 2\pi R.
\ena
This periodicity has two consequences:
\begin{enumerate}
\item
Momentum is quantized:
\bea
p=\frac{n}{R}, \quad
n \in \mathbb{Z},
\ena
\item
String can wrap the circle:
\bea
X(\s+2\pi) = X(\s) + 2\pi Rw, \quad
w \in \mathbb{Z}.
\ena
\end{enumerate}
In order to incorporate the effect of $w \neq 0$, it is necessary to
introduce separate mode expansions for left $(z)$ and right $(\bar z)$ movers:
\bea
X_L(z) &=& x_L - i\frac{\a'}{2} p_L \ln z + i\sqrt{\frac{\a'}{2}}\sum_{n\neq 0}
\frac{\a_n}{n} z^{-n}, \\
X_R(z) &=& x_R - i\frac{\a'}{2} p_R \ln \bar z + i\sqrt{\frac{\a'}{2}}
\sum_{n\neq 0} \frac{\tilde\a_n}{n} \bar z^{-n},
\ena
with
\bea
p_L = \frac{n}{R} + \frac{wR}{\a'}, \quad
p_R = \frac{n}{R} - \frac{wR}{\a'}.
\ena

Now Hamiltonian and momenta in this 2-dimensional theory are given by
\bea
H &=& \frac{1}{4\pi\a'}\int d\s [(\pa_t X^\mu)^2 +(\pa_\s X^\mu)^2] \nn
&=& \frac{\a'}{2}p^2 + \frac{\a'}{4}(p_L^2+p_R^2) + N + \tilde N, \\
P &=& \frac{1}{2\pi\a'}\int d\s \pa_t X^\mu \pa_\s X_\mu \nn
&=& \frac{\a'}{4}(p_L^2 - p_R^2) + N - \tilde N,
\ena
where $p_\mu$ is the uncompactified 9-dimensional momentum and
\bea
N \equiv \a_{-n}^\mu \a_{n\mu}, \quad
\tilde N \equiv \tilde\a_{-n}^\mu \tilde\a_{n\mu},
\ena
are the number operators of left- and right-movers.

Since the theory is a 2-dimensional gravity, we naturally have Hamiltonian
and momentum constraints (as relativists know quite well) on the physical
states, which yield the following conditions:
\bea
m^2 \equiv -p^2 &=& \frac{n^2}{R^2} + \frac{w^2R^2}{\a'{}^2}
 + \frac{2}{\a'}(N+\tilde N -2), \\
&& N-\tilde N + nw=0.
\ena
These determine the spectrum of the theory. Thus we find that the spectrum
is invariant under
\bea
R \leftrightarrow R' \equiv \frac{\a'}{R}, \quad
n \leftrightarrow w,
\ena
which means the symmetry of the spectrum in the theory under the transformation
\bea
p_L \to p_L, \quad
p_R \to -p_R.
\ena
Extending this to include nonzero modes as
\bea
X(z,\bar z) = X_L(z)+X_R(\bar z) \to
X'(z,\bar z) = X_L(z)-X_R(\bar z),
\ena
we find that this gives a symmetry of the theory, called
{\it T-duality}.~\cite{KY}

Under this transformation, world-sheet supersymmetry requires the
transformation
\bea
\tilde \psi'(\bar z) = -\tilde \psi(\bar z).
\ena
This then implies that the chirality (of right-mover) defined by
\bea
\tilde \Gamma= \tilde \psi_0^0 \psi_0^1 \cdots \psi_0^9,
\ena
is flipped. Namely if we start with type IIA theory, compactify the theory
to 9 dimensions and apply T-duality, the chirality of the fermions of
right-mover changes, transforming into type IIB, and vice versa. We thus
find that under T-duality type IIA and IIB theories are interchanged.

\section{T-duality of Open String}

\begin{figure}[htb]
\epsfysize=3cm \centerline{\epsfbox{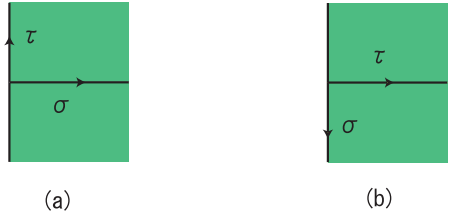}}
\caption{T-duality transformation}
\label{f1}
\end{figure}

Given that the closed string theories have T-duality, it is natural to
consider the same transformation in open string theories. The T-duality
transformation is actually the exchange of $\tau$ and $\s$ coordinates or
$\frac{\pi}{2}$ rotation on the worldsheet, as shown in Fig.~\ref{f1}

We then see that this transformation exchanges Neumann condition
$\pa_\s X^\mu =0$ and Dirichlet condition $\pa_\tau X^\mu=0$. This means that
even if we start with open string theories with Neumann boundary conditions,
we must have Dirichlet boundary conditions in T-dualized directions.
Thus there must be some objects which fix the open string end points.
Those are precisely the D-branes we mentioned before. When they have
spatial $p$-dimensional extension, they are called D$p$-branes.

It has been discovered~\cite{PO1} that D$p$-branes have this perturbative
description, and their important properties are as follows:
\begin{enumerate}
\item
They are BPS objects, namely they preserve part of supersymmetry
(typically 1/2).
\item
There is no force between parallel D$p$-branes computed by the diagram in
Fig.~\ref{f2}. This is due to the remaining supersymmetry and the exchange
of (NS, NS) and (R, R) bosons cancel.
\begin{center}
\begin{figure}[htb]
\epsfysize=3cm \centerline{\epsfbox{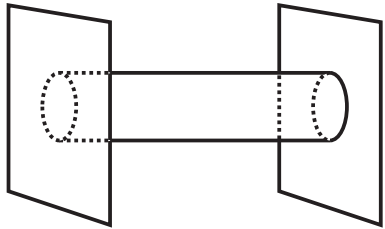}}
\caption{D-brane interaction}
\label{f2}
\end{figure}
\end{center}
\item
They are objects that carry RR charge ({\it i.e.} couple to RR forms present
in type II theories). Their tension is given by
\bea
\frac{1}{\a'{}^{(p+1)/2} g},
\ena
where $g$ is the string coupling constant, and their charges are quantized.
This is because $p$-brane and
$p' \equiv (6-p)$-brane is dual and Dirac quantization condition must hold
for the charges of these objects, just as the electrons and magnetic
monopoles. (Note that $p$-brane couples to $A_{p+1}$-form, whose field strength
$F_{p+2}$ is dual to $*F_{8-p}$, whose potential is $*A_{7-p}$-form which
couples to $(6-p)$-brane.) As we have discussed in \S~2, RR (potential)
forms of odd (even) rank exist in type IIA (IIB) theory, so IIA  (IIB)
contains D$p$-branes with even (odd) $p$.

The Dirac quantization condition on these charges is
\bea
\mu_p \mu_{6-p} = 2\pi n, \;\;
n \in \mathbb{Z}.
\label{qc}
\ena
In fact, the D$p$-brane charge is given by
\bea
\mu_p = (2\pi)^{\frac{7-2p}{2}}(\a')^{\frac{3-p}{2}},
\ena
consistent with the quantization condition~\p{qc}.
This will be important in studying black hole entropy.
\item
Open strings can attach to D-branes. This means that the effective theory
on the D-branes is a gauge theory.~\cite{WI2}
\end{enumerate}
All of the above properties turned out to play very important roles
in the understanding of nonperturbative properties of superstring theories.

\section{T-duality of Unoriented String}

Consider, in closed string, worldsheet parity transformation
\bea
\Omega: \s \to 2\pi -\s,
\ena
and require that all the states in the theory obey
\bea
\Omega =+1.
\ena
Physically this implies that all the states in the theory are invariant
under the orientation reversal, leaving a theory of unoriented strings.
If we consider T-duality in a theory with this requirement,
the dual coordinate $X'{}^m(z,\bar z)=X_L^m(z)-X_R^m(\bar z)$ transforms as
\bea
X'{}^m(z, \bar z) \leftrightarrow - X'{}^m(z,\bar z),
\ena
whereas the original coordinates remain the same.
The symmetry in the dual picture is then
\bea
\hs{-10}
\mbox{(parity transf. }\s \to 2\pi -\s)
\times(\mbox{reflection in $m$ direction } X'{}^m \to - X'{}^m).
\ena
This requirement defines unoriented string. The target space is not a circle
but a half line, which is $S^1/{\mathbb Z}_2$, and is called {\it orientifold}.

As a simple example of orientifold, consider a circular coordinate
$-\pi R \leq x \leq \pi R$. The above ${\mathbb Z}_2$ identification makes
the space
\bea
0 \leq x \leq \pi R.
\ena
and produces fixed planes at $x=0,\pi R$, called {\it orientifold fixed
planes}. Note that in this construction these are rigid fixed planes
and not dynamical ones unlike D-branes.

Let us look at the theory in various aspects.
Close to the orientifold, we have unoriented open strings
as well as closed strings. In this way {\it type I theory may be regarded as
type IIB theory on orientifold.}

Far from these fixed planes, the theory is just like the original oriented
IIB string, and the unorientedness is taken care by the presence of the fixed
planes which fix the motion of the strings at mirror points.

We are now ready to discuss the M-theory conjecture.

\section{Strong coupling Limit of Type II Superstring}

In type IIA theory, we have learned that there exist D$p$-branes for even
$p$. In particular, consider D0-branes. Their properties of particular
importance here include:
\begin{enumerate}
\item
The tension or energy of D0-branes is $\frac{1}{\sqrt{\a'}g}$.
\item
They are BPS, so that the energy of $n$ bound states is
$\frac{n}{\sqrt{\a'}g}, (n=1,2, \cdots)$ without binding energy.
\item
Their BPS property (or supersymmetry representation theory) means that this
energy is not changed by quantum corrections.
\end{enumerate}

In perturbation theory, D0-branes can be neglected because they are heavy
with mass inversely proportional to the coupling constant.
In the strong coupling, however, they are light and the low-energy
effective theory is significantly modified. Not only that, we have a huge
tower of light massive states as given in item 2 above.

How can we understand such an infinite tower of light massive states?

The proposed answer is that they are just the Kaluza-Klein(KK)-modes from
11 dimensions! This is the first signal of the M-theory.
There are also various other evidences for this proposal in addition to
the fact that other BPS D-branes can be understood similarly.

To get more concrete relation between the 11-dimensional M-theory and
superstrings, consider massless fields in type IIA theory:
\bea
&& \mbox{(NS, NS): dilaton } \phi,
\mbox{ metric } g_{\mu\nu}, \mbox{ antisymmetric } B_{\mu\nu}, \nn
&& \mbox{(R, R): forms } A_1, A_3.
\ena
The (bosonic part of) the low-energy effective action (IIA supergravity) is
\bea
S_{\rm NS-NS} &=& \frac1{2\kappa_{10}^2} \int d^{10}x \sqrt{-g} e^{-2\phi}
\Big( R + 4 (\nabla \phi)^2 -\frac{1}{2} H^2 \Big),\nn
S_{R-R} &=& -\frac{1}{4\kappa_{10}^2}\int d^{10}x\sqrt{-g}\Big(F_2^2
+F_4'{}^2\Big) -\frac1{4\kappa_{10}^2}\int F_4 \wedge F_4 \wedge B,
\label{iiasugra}
\ena
where $2\kappa_{10}^2= 16\pi G_{10}$ is the 10-dimensional Newton constant and
\bea
H=dB,\;\;
F_2 = dA_1,\;\;
F_4 = dA_3,\;\;
F_4' = dA_3+A_1 \wedge H,
\ena
and the suffices indicate the ranks of the forms.
The algebra of the two supercharges of opposite chirality $Q, Q'$
roughly takes the form
\bea
&& \{Q,Q\} \sim \{Q', Q'\} \sim P, \nn
&& \{Q, Q' \} \sim W \ \ (\cdots \mbox{ central charge}).
\label{cc}
\ena
The D-branes belong to the representation of this algebra with maximum
central charges.

We can understand the origin of the central charges in \p{cc} as follows:
When the 11-dimensional theory is compactified on ${\mathbb R}^{10} \times
S^1$, the fields in the original theory produce the gauge fields and central
charges as
\bea
g_{\mu,11} \Rightarrow A_1, \quad
p_{11} \Rightarrow W.
\ena
Namely the central charge is nothing but the 11-th momentum and its gauge
field is the RR-form originating from the 11-dimensional metric.
This means that the RR charged objects are actually KK modes. Also their
mass is given by $\frac{n}{R_{11}}, n\in \mathbb{Z}$, which is of the
same form as the mass of D0-branes. The 10-dimensional $N=2$ supersymmetry
also matches with 11-dimensional supersymmetry.

Comparison of the low-energy actions (massless) tells us that we should
write the 11-dimensional metric in terms of the type IIA fields as
\bea
ds_{11}^2 &=& e^{-\frac{2}{3}\phi}g_{\mu\nu}dx^\mu dx^\nu
+ e^{\frac{4}{3}\phi}(dy +A_\mu dx^\mu)^2.
\label{11metric}
\ena
This yields in fact the effective action~\p{iiasugra} from the 11-dimensional
action
\bea
S_{11} = \frac{1}{2\kappa_{11}^2}\int d^{11}x \sqrt{-G}\left( R
-\frac{1}{2}|F_4|^2 \right)-\frac{1}{12\kappa_{11}^2}\int A_3\wedge F_4
\wedge F_4,
\ena
where $\kappa_{11}^2 (= 2\pi R_{11}\kappa_{10}^2)$ is the 11-dimensional
Newton constant. The metric~\p{11metric} shows that the 11-dimensional
radius is given by
\bea
R_{11}= e^{\frac{2}{3}\phi} = g^{\frac23}.
\ena
These results resolve several questions:
\begin{enumerate}
\item
Perturbation in the string coupling $g$ is equivalent to the expansion
around $R_{11}=0$.
This is the reason why string perturbation cannot see 11-th dimension!!
\item
Masses of KK modes are given by
\bea
\frac{e^{-\frac13\phi}}{R_{11}} \sim \frac{1}{g},
\ena
which in fact matches with those of D0-branes. They are also BPS states with
RR-charge, consistent with D0-branes.
\end{enumerate}

In the strong coupling limit $g \to \infty, R_{11}\to \infty$ and the theory
is an 11-dimensional supersymmetric theory with gravity. The low-energy
effective theory (containing only massless degrees) must be the 11-dimensional
supergravity which is the unique theory with this property.

We have thus learned that the strong coupling limit of type IIA theory is
the M-theory. What about the strong coupling limit of type IIB theory?

Consider the metrics for IIA and IIB compactified on $S^1$ of radii $R_A$
and $R_B$, respectively. The T-duality relation of these theories implies
that the metrics are written as
\bea
{\rm IIB}/R_{\rm B} &:& ds^2 = \cdots + g_{99} (dx^9)^2, \nn
{\rm IIA}/R_{\rm A} &:& ds^2 = \cdots + \frac{1}{g_{99}} (dx^9)^2, \quad
g_{99} = e^{2(\phi_B-\phi_A)}, \nn
&& R_A = e^{\phi_A - \phi_B}, \quad
R_B = e^{-\phi_A + \phi_B},
\label{iiabmet}
\ena
where $\phi_{A,B}$ are the dilatons in each theory. When these theories
are regarded as M-theory compactified on $T^2$ of radii $R_{10},R_{11}$,
the relation~\p{11metric}
\bea
ds_{11}^2 = e^{-\frac{2}{3}\phi_A}ds_{10}^2 + e^{\frac{4}{3}\phi_A}
 (dy +A_\mu dx^\mu)^2,
\ena
yields $R_{10}= e^{-\frac{1}{3}\phi_A} R_A$. A simple manipulation of
metrics and radii using relations given in \p{iiabmet} gives the type IIB
coupling as
\bea
g^{(B)} (\equiv e^{\phi_B}) = \frac{R_{11}}{R_{10}}.
\ena
Since $R_{10}$ and $R_{11}$ are completely on the same footing, we thus find
that the IIB theory is invariant under
\bea
g^{(B)} \to \frac{1}{g^{(B)}}.
\label{s}
\ena
Consequently the strong coupling limit of IIB is itself!
More generally, it is known the effective type IIB supergravity is invariant
under $SL(2,{\mathbb Z}_2)$ transformation
\bea
\tau \to \frac{a\tau +b}{c\tau +d}; \;\;
\tau \equiv l +ie^{-\phi_B}, \quad
a,b,c,d \mbox{: integers}, \;\;
(ad-bc=1),
\label{sl2}
\ena
where $l$ is an RR scalar field. Note that \p{sl2} reproduces \p{s} for
$l=0$ and $a=d=0, b=-c=1$. It is believed that the complete type IIB
superstring (not only the massless sector) has this invariance.
There is also another evidence supporting this conjecture.

\begin{center}
\begin{figure}[htb]
\epsfysize=2cm \centerline{\epsfbox{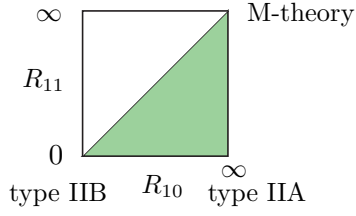}}
\begin{picture}(10,-5)
\put(150,40){\small $R_{11}$}
\put(195,0){\small $R_{10}$}
\put(160,11){0}
\put(155,65){\small $\infty$}
\put(225,6){\small $\infty$}
\put(145,-3){\small type IIB}
\put(220,-3){\small type IIA}
\put(235,65){\small M-theory}
\end{picture}
\caption{IIB or not IIB}
\label{f3}
\end{figure}
\end{center}

Thus we have learned
\vs{2}
\begin{center}
\begin{tabular}{|ccccccc|}
\hline
IIB & $\Leftrightarrow$ & IIB & $\Leftrightarrow$ & IIA & $\to$ &
11-dim. M-theory \\
 & strong & & T-duality & & strong & \\
 & coupling & & $R \leftrightarrow \frac{\a'}{R}$ & & coupling &  \\
 & & & & & {$R_{11} \to \infty$} & \\
\hline
\end{tabular}
\end{center}
\vs{2}
The overall picture is like in Fig.~\ref{f3}.

\section{$SO(32)$ Type I and Heterotic Strings}

Once the evidence for the existence of the unifying M-theory was discovered,
it was quickly accepted and duality for other superstring theories was
easily found. In the following few sections, we will briefly summarize the
evidence of the dualities of all superstring theories.

The first is the duality between $SO(32)$ type I and heterotic
strings.~\cite{PW} We note that $SO(32)$ type I theory possesses $D=10, N=1$
supersymmetry. This theory contains open and closed strings in the weak
coupling picture. The question we address here is what happens in the strong
coupling limit of this theory.

We first note that the maximum spin in the massive representations of
$D=10, N=1$ supersymmetry algebra is 2. It is impossible that these massive
spin 2 particles become massless in the strong coupling limit and the theory
is promoted to 11-dimensional theory, because 11-dimensional theory has
bigger ($N=2$) supersymmetry and no gauge particles.

The only possibility would be then to reduce to again 10-dimensional theory,
which cannot be itself because there is no $SL(2,{\mathbb Z}_2)$-like
symmetry in type I theory (contrary to IIB). It turns out that the theory
we reach is the heterotic $SO(32)$ theory in the strong coupling limit.
\vs{2}
\begin{center}
\begin{tabular}{|ccc|}
\hline
$SO(32)$ type I theory & $\Leftrightarrow$ & heterotic $SO(32)$ theory \\
 & S-dual & \\
\hline
\end{tabular}
\end{center}

\vs{2}
Let us check this relation by comparing the low-energy effective theories.
The type I effective theory is given by
\bea
S_I &=& \frac{1}{2\kappa_{10}^2}\int d^{10}x \sqrt{-G}\left\{
e^{-2\phi}(R+4(\pa_\mu \phi)^2) -\frac{1}{2}|\tilde F_3|^2 \right\} \nn
&& -\frac{1}{2g_{10}^2}\int d^{10}x\sqrt{-G} e^{-\phi}{\rm Tr}|F_2|^2,
\ena
where $\tilde F_3 = dC_2 -\frac{\kappa_{10}^2}{g_{10}^2} \omega_3,
\omega_3= {\rm Tr}(A_1\wedge dA_1 -\frac{2}{3}i A_1^3)$, and $g_{10}$ is
the SYM coupling constant.
On the other hand, the $SO(32)$ heterotic effective theory is given by
\bea
S_H &=& \frac{1}{2\kappa_{10}^2}\int d^{10}x \sqrt{-G} e^{-2\phi}
\Big\{ R+4(\pa_\mu \phi)^2 -\frac{1}{2}|\tilde H_3|^2
-\frac{\kappa_{10}^2}{g_{10}^2}{\rm Tr}|F_2|^2\Big\},
\ena
where $\tilde H_3 = dB_2 -\frac{\kappa_{10}^2}{g_{10}^2} \omega_3.$
It is easy to see that these two effective theories are related by
\bea
\begin{array}{ll}
G_{I\mu\nu} = e^{-\phi_H}G_{H\mu\nu}, & \phi_I = -\phi_H, \\
\tilde F_{I3}= \tilde H_{H3}, & A_{I1}=A_{H1},
\end{array}
\ena
in agreement with the above assertion.

\section{Two Heterotic Strings}

Using the similar argument, it is not difficult to show that there is a
relation between two heterotic string theories with $SO(32)$ and
$E_8\times E_8$ symmetries under T-duality.~\cite{GI} In order to show this,
we have to first break the gauge symmetries to the common $SO(16)\times SO(16)$
symmetry by introducing Wilson lines in the moduli space of the theories,
and then use the properties of the even self-dual lattice in the moduli
space to argue the equivalence of the resulting theories. In this way it has
been shown that these theories are transformed into each other under
T-duality transformation. Schematically this is written as
\begin{center}
\begin{tabular}{|ccc|}
\hline
$SO(32)$ heterotic & $\Leftrightarrow$ & $E_8 \times E_8$ heterotic \\
$\downarrow$ & {\normalsize $S^1$ with} & $\downarrow$ \\
& {\normalsize Wilson line} & \\
$SO(16)\times SO(16)$ heterotic & $\Leftrightarrow$ & $SO(16)\times SO(16)$
 heterotic
\\
& T-dual & \\
\hline
\end{tabular}
\end{center}
and hence
\begin{center}
\begin{tabular}{|ccc|}
\hline
$SO(32)$ heterotic & $\Leftrightarrow$ & $E_8 \times E_8$ heterotic \\
& T-dual & \\
\hline
\end{tabular}
\end{center}

\section{ String Web --- Overall picture}

What we have learned so far can be summarized in the following diagram:
\begin{center}
\begin{tabular}{|cccccccc|}
\hline
11D & M-theory & & & & & & \\
 & $|S^1$ & & & & & & \\
10D & IIA & $\Leftrightarrow$ & IIB & & & & \\
& & T & \hs{5} $|S^1/{\mathbb Z}_2$ & & & & \\
9D & & & type I & $\Leftrightarrow$ & $SO(32)$ heterotic & $\Leftrightarrow$
& $E_8\times E_8$ heterotic \\
& & & & S & & T & \\
\hline
\end{tabular}
\end{center}
\vs{2}
This already indicates that all superstrings are dual.
Using this relation and the connection between IIB and I superstrings
discussed in \S~6, we start with M-theory compactified on $S^1\times S^1
/{\mathbb Z}_2$ and end with $E_8\times E_8$ heterotic theory on $S^1$
by the following route:
\begin{center}
\begin{tabular}{|ccccc|}
\hline
M/$S^1\times S^1/{\mathbb Z}_2$ & $\to$ & IIA/$S^1/{\mathbb Z}_2$ & $\to$ &
IIB/$S^1/{\mathbb Z}_2$ \\
& S & & T & $\downarrow$ \\
$E_8\times E_8$ heterotic/$S^1$ & $\leftarrow$ & $SO(32)$ heterotic/$S^1$
& $\leftarrow$ & I/$S^1$ \\
& T & & S & \\
\hline
\end{tabular}
\end{center}
Comparing the both ends, we arrive at the conjecture by Ho\v{r}ava and
Witten~\cite{HW}
\bea
\mbox{M/$(S^1/{\mathbb Z}_2)$ = $E_8\times E_8$ heterotic}.
\label{mh}
\ena
As we have discussed in \S~6, $S^1/{\mathbb Z}_2$ is an orientifold, a line
with fixed planes on both ends.

There are several supporting evidences for this conjecture:
First, the low-energy limit of the M-theory is the 11-dimensional supergravity.
The above compactification~\p{mh} is consistent with the 11-dimensional
supergravity because it is invariant under ${\mathbb Z}_2$ if $A^{(3)} \to
-A^{(3)}$ so that we can consider this compactification.

Second, it is consistent with supersymmetry.
M/${\mathbb R}^{10}\times S^1$ is invariant under supersymmetry with arbitrary
$\e$ (32 components), namely this gives $N=2$ in 10 dimensions.
${\mathbb Z}_2$ action kills half of the supersymmetry, because unbroken
supersymmetry is determined by the condition $\Gamma^{11}\e =\e$. This also
means that $\e$ is chiral in 10 dimensions, consistent with the supersymmetry
of the heterotic string!
It follows that M/${\mathbb R}^{10}\times S^1/{\mathbb Z}_2$ reduces
to 10-dimensional Poincar\'{e}-invariant theory with one chiral supersymmetry.
There are three candidates satisfying this criterion:
$E_8\times E_8$ heterotic,
$SO(32)$ heterotic, and
$SO(32)$ type I theories.
Which is the one we are looking for?

The answer is given by the following considerations, all of which point to
$E_8 \times E_8$ heterotic theory.

{\it (i) gravitational anomaly}\\
The effective action must be invariant under diffeomorphism.
On smooth 11-dimensional manifold, the theory is anomaly free. However,
when it is compactified on orientifold, 11-dimensional Rarita-Schwinger
field produces not only infinitely many massive fields (which are anomaly free)
but also massless chiral 10-dimensional gravitini which may produce anomaly.

The absence of anomalies in 11 dimensions implies that the anomalies do not
exist at smooth point in ${\mathbb R}^{10}\times S^1/{\mathbb Z}_2$.
It then follows that the possible anomalies are sum of delta functions on the
fixed hyperplanes at $x^{11}=0,\pi R$.
By symmetry, the form of anomalies at $x^{11}=0,\pi R$ must be the same.
Since the anomaly is not zero, there must be additional massless modes that
propagate only on the fixed planes and cancel those from gravitini.
They must be 10-dimensional fields, and the only candidates are the
10-dimensional vector multiplets.
In 10 dimensions, standard anomaly can be canceled only by 496 vector
multiplets, which are divided equally between the two fixed hyperplanes.
Consequently 248 vector multiplets must exist on each hyperplane, implying
that they must be $E_8\times E_8$ gauge particles. $SO(32)$ is impossible
because that would require all the vector multiplets on one hyperplane.

{\it (ii) Strong coupling behavior}\\
If M-theory on ${\mathbb R}^{10}\times S^1/{\mathbb Z}_2$ of radius $R$
is equivalent to $E_8\times E_8$ heterotic string with coupling constant $g$,
we have $R=g^{2/3}$. This means that when $R$ is small, string picture is
a good description. When $R$ is large, on the other hand, supergravity
description is good. It follows that the relation can be well studied
in the supergravity approximation.

Now we already know (in \S~8) that the strong coupling limit of type I
superstring in 10 dimensions is again a 10-dimensional weakly coupled $SO(32)$
heterotic string, and hence these two are not related to 11-dimensional
supergravity in the strong coupling or large $R$ limit. The only possibility
left is then that the orientifold theory in the large $R$ limit is related to
$E_8\times E_8$ heterotic string.

Thus all evidences support the conjecture that the M-theory compactified on
$S^1/{\mathbb Z}_2$ is the $E_8 \times E_8$ heterotic theory.

In this way all consistent superstrings are related with each other by
S-duality or T-duality. This completes the whole picture of the string web.

\section{ D-branes in Type II Supergravity and Black Holes}

We now go on to describe the same D-branes as the classical soliton solutions
in the low-energy effective theory (supergravity). This allows us to
understand the geometry of the space-time produced by D-branes, and
gives black hole interpretation of these solutions.

The massless degrees of freedom in the theories are
gravity + dilaton + antisymmtric tensors (plus fermions which are irrelevant
to our following discussions). The effective actions are uniquely determined
by 10- or 11-dimensional supersymmetry.
They allow various brane solutions as solitons.
Most important are the D-branes. The previous description of
D-branes is in terms of perturbative string picture. Here the D-branes are
identified as the classical solitonic solutions in supergravity.

The bases to identify these classical solutions as D-branes are:
\begin{enumerate}
\item
They are spatially $p$-dimensional extended BPS objects with 1/2 supersymmetry.
\item
They carry RR charges.
\end{enumerate}

The important assumption is that the physical contents of the theory do not
change if coupling constant is changed
because of the remaining supersymmetry (BPS representation).
This means that there are two ways of description of the same
objects -- D-branes -- by open and closed string degrees of freedom.

{\it Example:} Let us consider 2-brane solution in $D=11$
\bea
ds^2 &=& H^{1/3}[ H^{-1}(-dt^2 +dy_1^2 + dy_2^2) +dr^2 + r^2 d\Omega_7^2],\\
&& A_{\mu\nu\lambda}=\e_{\mu\nu\lambda} H^{-1},\;
H \equiv 1+ \frac{k}{r^6}.
\ena
We regard the 8-dimensional space described by $r$ and angular coordinates
as our space-time. The metric is invariant in $y_1, y_2$ directions, and is
asymptotically flat in other directions. This means that there is some object
extended in $y_1,y_2$, which deserves the name of 2-brane or membrane.
\begin{figure}[htbp]
\epsfysize=3cm \centerline{\epsfbox{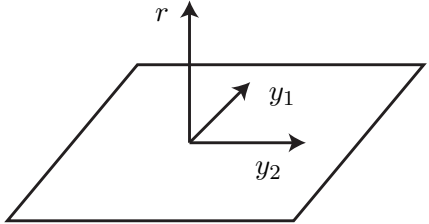}}
\begin{picture}(0,0)
\put(177,90){$r$}
\put(220,60){$y_1$}
\put(215,32){$y_2$}
\end{picture}
\caption{Membrane}
\label{f4}
\end{figure}

\begin{figure}[htbp]
\epsfysize=9cm \centerline{\epsfbox{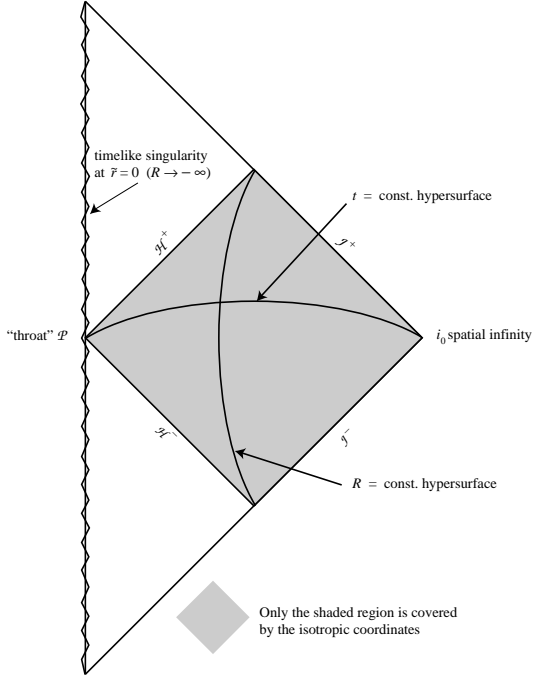}}
\caption{Carter-Penrose diagram for M2-brane.~\cite{STE} Here
$R=(1-k/\tilde r^6)^{1/3}$.}
\label{f5}
\end{figure}

Now take the Schwarzschild-type coordinates $r=(\tilde r^6 - k)^{1/6}$, and
the metric is written as
\bea
\hs{-10}
ds^2 = (1-k/\tilde r^6)^{2/3}(-dt^2 +dy_1^2 + dy_2^2)
 +(1-k/\tilde r^6)^{-2}d\tilde r^2 + \tilde r^2 d\Omega_7^2.
\ena
If $y_1$ and $y_2$ are compactified, we see that this describes a black hole
geometry with a horizon at $\tilde r=k^{1/6}$, but a time-like and space-like
vectors are not interchanged upon crossing it. The causal structure of
this space-time is depicted in Fig.~\ref{f5}.
We find that the global structure is similar to the Reissner-Nordstrom
solution.

Similar extended brane solutions in superstring theories are summarized
in the following table:
\begin{center}
\begin{tabular}{l|l}
type IIA & type IIB \\
\hline
fundamental string & fundamental string \\
D$p$-branes ($p$: even) & D$p$-branes ($p$: odd) \\
NS5-brane & NS5-brane \\
KK-wave & KK-wave
\end{tabular}
\end{center}
More general solutions can be constructed by combining these
solutions.~\cite{PT,T,AEH,NO4} The construction rules are called
intersection rules.
Combining these according to the intersection rules, it is possible to make
solutions that can be interpreted as $D=4,5$ black holes.

{\it Example:}
Type IIB D5-D1-KK-wave intersecting solution is given by
\bea
&&\hs{-15}
 ds^2 = H_1^{-\frac34} H_5^{-\frac14}[-(1-r_0^2/r^2) dt^2 + dx_9^2
+(r_0^2/r^2)(\cosh \s dt + \sinh \s dx_9)^2] \nn
&& \hs{-10}+ H_1^{\frac14} H_5^{-\frac14}[dx_5^2 + dx_6^2 + dx_7^2+ dx_8^2]
+ H_1^{\frac14} H_5^{\frac34}[(1-r_0^2/r^2)^{-1}dr^2 + dr^2 \Omega_3^2],
\ena
where $r^2=x_1^2 + \cdots +x_4^2, H_i = 1+\frac{r_0^2}{r^2}\sinh \a_i$,
$x_5, \cdots,x_9$ are compactified on a torus $T^5$ of radii $R_5, \cdots,R_9$,
giving $D=5$ black hole.
\begin{figure}[htbp]
\begin{center}
\epsfysize=2cm \centerline{\epsfbox{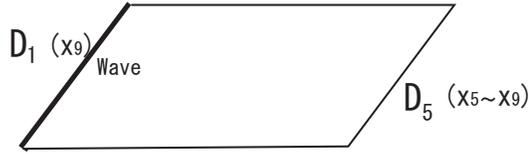}}
\end{center}
\caption{D1-D5-Wave intersecting solution}
\label{f6}
\end{figure}
Remember that the D-brane charges are quantized. This leads to
the result that the Bekenstein-Hawking entropy
\bea
S =\frac{\rm area}{4} = 2\pi \sqrt{Q_1 Q_5 N},
\label{ent}
\ena
is quantized, where $Q_1, Q_5$ and $N$ are the (integer) numbers of D1-,
D5-branes and momentum!! This is already quite a nontrivial result which
is obtained from the brane realization of black hole solutions.

Now we are going to use another description of this black hole or D-brane
solutions in terms of perturbative string picture to give explanation of the
entropy~\p{ent} by statistical mechanics.
The existence of the D-branes is probed by open string.
This means that if there is only one D-brane, $U(1)$ gauge theory is realized
on the D-brane. If there are two separate D-branes, then $U(1)\times U(1)$
plus open strings between the 2 D-branes exist. When these 2 D-branes are
on top of each other, gauge symmetry is enhanced to $U(2)$. Thus
$Q_5$ D-branes at the same position give rise to $U(Q_5)$ gauge theory on
the world-volume!

Using this description, we can count the degrees of freedom living on the
above solution. The system can be regarded as gas of
$N_B=N_F=4 Q_1 Q_5$ massless particles with energy $E=N/R_9$ in a space of
length $L=2\pi R_9$, giving the entropy precisely agreeing with \p{ent}.
We thus get precise agreement of the black hole
entropy, calculated from Bekenstein-Hawking formula and from this string
picture.~\cite{SV,CM}

This gives the statistical origin of the black hole entropy.
The upshot is that the information is stored on the brane or horizon, and
the counting of the degrees of freedom on them gives the statistical
mechanical entropy which coincide with the Bekenstein-Hawking entropy~\p{ent}.
This is a kind of holographic principle,
meaning that information is stored on the boundary.

This is also one of the manifestations of the equivalent descriptions
(duality) of the theory in terms of open and closed string degrees of freedom.
We are now going to describe another manifestation of this duality
under the name of AdS/CFT correspondence.

\section{ AdS/CFT Correspondence}

Let us consider S-channel and T-channel duality depicted in the following
diagrams in Fig.~\ref{f7}:
\begin{figure}[htbp]
\begin{center}
\epsfysize=2cm \centerline{\epsfbox{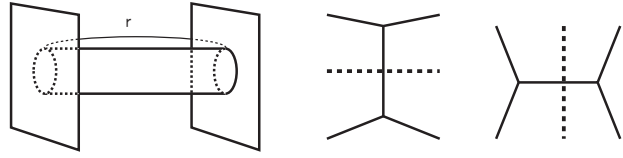}}
\end{center}
\caption{S- and T-channel duality}
\label{f7}
\end{figure}
\\
The claim is that the latter two diagrams give the same amplitudes in string
theories. When applied to the first diagram, this shows the remarkable
property of the string theory that the open string one loop diagram is
equivalent to that of closed string exchange.
This equivalence is valid after sum over all modes.
When the distance $r$ of closed string propagation is small, this process
can be better described by open strings, or in terms of SYM
theories. However, when the length $r$ of open string is large, it is better
described by closed strings, or in terms of supergravity.

This correspondence or duality of the two descriptions may be best explained
by the following example of D3-branes:
\bea
ds^2 = H^{-\frac{1}{2}}(-dt^2 + dy_1^2 + dy_2^2 + dy_3^2) +H^\frac{1}{2}(dr^2
+r^2 d\Omega_5^2),
\label{d3metric}
\ena
Near branes ($r \sim 0$), description by open strings is good and the theory
is well described by $D=4, N=4$ SYM theory. This theory possess superconformal
symmetry including $SO(2,4)$.
On the other hand, the same theory may be expected to be well described by
the near-horizon geometry in supergravity solution~\p{d3metric} for certain
region of coupling constant. Near the horizon, \p{d3metric} is approximated as
\bea
ds^2 \sim \a'\Big[\frac{u^2}{\sqrt{4\pi gN}} dx_4^2
 +\sqrt{4\pi gN}\Big(\frac{du^2}{u^2}+ d\Omega_5^2\Big) \Big],
\label{ads5}
\ena
where $u\simeq r/\a'$ is fixed. (Here $u$ gives the energy scale we are
looking at the theory, and fixing $u$ means that we are keeping the
stretched open string mass finite.) This is a space of direct product of
AdS$_5 \times S^5$ with the {\it same superconformal symmetry} including
$SO(2,4) \times SO(6)$.

It has been proposed that the above $D=4,N=4$ superconformal theory is
well described by this AdS solution~\p{ads5}.~\cite{MAL}
The evidences that the above two descriptions are valid include:
\begin{enumerate}
\item
Agreement of symmetries, as shown above.
\item
Agreement of spectrum.
\item
Agreement of operator algebra.
\end{enumerate}
This is what is called AdS/CFT correspondence or duality, meaning that there
are different descriptions of the same object valid for different regions
of coupling constant etc.
This is closely related to 't Hooft's old idea that the large $N$ gauge theory
(open string) is related to string (closed string) in the confining phase.

Note that the curvature for the metric~\p{ads5} is proportional to
$\frac{1}{\sqrt{4\pi gN}}$, so supergravity description is good in the
large $gN$ but small $g$ limit, and hence in the large $N$ limit.
Otherwise we have to consider full string theory.

The rules for practical calculations are formulated in refs.~\citen{WI3,GKP}
and they are based  on the following observations:
In AdS background, the bulk field $\phi$ is completely determined
by its field equation and given boundary values $\phi(\Omega)$.
Hence the action gives the generating functional of the correlation
functions for the local operators on the boundary which couple to
the boundary value of the fields.
\bea
e^{iS_{eff}(\phi_i)} = \lan T e^{i\int_B \phi_{b,i}{\cal O}^i}\ran
\ena
In the next section, we give greybody factors for BTZ black holes
by using this idea.~\cite{BSS,MOZ,O}

Before going into details, let us summarize properties of AdS$_{p+2}$ space,
which is a space with $R_{\mu\nu}=\frac{\Lambda}{D-2}g_{\mu\nu}$ with
$D>2, \Lambda <0$. Namely AdS$_{p+2}$ is a space with negative cosmological
constant. A convenient description of the space is to use the embedding in
flat $(p+3)$-dimensional space-time:
\bea
-y_{-1}^2 -y_0^2 + y_1^2 +\cdots + y_{p+1}^2 = -R^2,
\ena
with $SO(2,p+1)$ isometry.
\bea
ds^2 = -dy_{-1}^2 -dy_0^2 + dy_1^2 +\cdots + dy_{p+1}^2,
\ena
Put
\bea
r=y_{-1}+y_{p+1},\;\;\;
v=y_{-1}-y_{p+1},\;\;\;
x_i = (R/r)y_1,
\ena
and we get
\bea
ds^2=\frac{r^2}{R^2}dx^2 + \frac{R^2}{r^2} dr^2
= R^2(U^2 dx^2 + \frac{dr^2}{U^2}) ,\;\;\;
U=\frac{r}{R^2}.
\ena
For $p=3$, this has the same symmetry as $N=4,D=4$ SCFT with $SO(2,4)$,
leading to the conjecture that the strong coupling SYM is equivalent to
the weak coupling supergravity.

Finally let us summarize the known examples of AdS$_{p+2}$/CFT$_{p+1}$
correspondence in the following table:
\begin{center}
\begin{tabular}{l|l|l}
\hline
M-theory & M2 ($p=2$) & AdS$_4$/CFT$_3$ \\
& M5 ($p=5$) & AdS$_7$/CFT$_6$ \\
\hline
IIB & D3 & AdS$_5$/CFT$_4$ \\
 & D1 + D5 on T$^4$ & AdS$_3$/CFT$_2$ \\
\hline
\end{tabular}
\end{center}
\vs{2}
The last one is the AdS$_3$/CFT$_2$ correspondence best checked.

{\it Related results and future directions:}
\begin{itemize}
\item
The entropy for BTZ (and higher-dim.) black hole can be computed using the
central charge of the Virasoro algebra realized on the boundary geometry
without recourse to supersymmetry.~\cite{STR,ISA}
\item
The properties of gravity, in particular the evolution of Hawking
radiation and the information loss, can be studied by well-behaved, unitary
conformal field theory. If true, this suggests that the problem of information
loss is actually not present.
\item
It may be even possible to formulate ``string theory'' in terms of
field theory. A partial realization of this idea is what is called
Matrix theory.~\cite{BFSS}
\end{itemize}

We now discuss two applications of AdS/CFT correspondence in the following
two sections.

\section{ Greybody Factors for BTZ Black Hole}

Consider the 3-dimensional BTZ black hole. When embedded in string theory,
this can be related to other 5D and 4D black holes, whose metric is
\bea
ds_{10}^2 &=& - \frac{(\rho^2 - \rho_-^2)(\rho^2 - \rho_+^2)}
 {l^2 \rho^2} dt^2 + \rho^2 \left(d\varphi
 - \frac{\rho_+ \rho_-}{l \rho^2} dt \right)^2 \nn
&& + \frac{l^2 \rho^2}{(\rho^2 - \rho_+^2)(\rho^2 - \rho_-^2)}d\rho^2
 + l^2 d\Omega_3^2 + dy_2^2 + \cdots + dy_4^2.
\ena
The first part is the metric for BTZ black hole!
Thus the entropies of 5D and 4D black holes may be related to those of
BTZ black holes.\cite{BSS,MOZ,O} We now claim that not only the entropy of
black hole but also absorption cross sections or greybody factors can
be examined by going to AdS$_3$ and using the AdS/CFT correspondence.

The basic idea is the following: The greybody factors can be evaluated for
BTZ black holes through the discontinuity of the two-point correlation
functions (optical theorem) evaluated from AdS/CFT correspondence.

In AdS space in the Poincar\'{e} coordinate
\bea
ds^2 = \frac{l^2}{y^2} ( dy^2 + dw_+ dw_-),
\ena
we consider massive scalar field with mass $m$:
\bea
S(\phi)= \frac12 \int dy dw_+ dw_- \sqrt{g} \left(
g^{\mu\nu} \pa_\mu \phi \pa_\nu \phi + m^2 \phi^2 \right),
\ena
which has a solution with the behavior
\bea
\phi(y,w_+,w_-) \to y^{2h_-} \phi_0 (w_+,w_-),
\ena
for $y \to 0$ (boundary), with the dimension of the boundary value
$\phi_0 (w_+,w_-)$ and
\bea
h_\pm = \frac12 (1 \pm \sqrt{1+m^2}).
\ena

The main steps in our calculations are:
\begin{enumerate}
\item
Evaluate the two-point function by AdS/CFT correspondence.
\item
Obtain the greybody factors in BTZ black holes from the
discontinuity of the two-point correlation function (optical theorem).
\end{enumerate}
After this procedure, we find~\cite{MOZ}
\bea
\sigma_{abs} &=&
 \frac{\pi}{\omega} \int dt \int_{-\infty}^{\infty}d\varphi e^{ip\cdot x}
 [ G(t-i\e,\varphi) - G(t+i\e, \varphi)] \nn
&=& \frac{2h_+(2h_+-1) (2\pi T_+ l)^{2h_+-1}(2\pi T_- l)^{2h_+-1}
 \sinh\left( \frac{\omega}{2 T_H} \right)}{\omega \Gamma^2(2h_+)} \nn
&& \times \left| \Gamma\left(h_+ + i\frac{\omega}{4\pi T_+} \right)
 \Gamma\left(h_+ + i\frac{\omega}{4\pi T_-} \right) \right|^2,
\ena
where
\bea
\frac{2}{T_H} = \frac{1}{T_+} + \frac{1}{T_-},\quad
T_\pm \equiv \frac{\rho_+ \mp \rho_-}{2\pi}.
\ena
We can read off the decay rate for massless scalar field from this result:
\bea
\Gamma &=& \frac{\sigma_{abs}(h_+=1)}{e^{\omega/T_H}-1} \nn
&=& \frac{\omega \pi^2 l^2}{(e^{\omega/2T_+}-1)(e^{\omega/2T_-}-1)},
\ena
consistent with the semiclassical gravity calculations, giving another
evidence of the duality.

The initial state of BTZ black holes was described by Poincar\'{e}
vacuum, and the calculation involves a nonlinear coordinate transformation,
which induces the Bogoliubov transformation on the operators. This is the
origin of the thermal factor in the above.

It is possible to use the above results and methods to evaluate similar
quantities in 4D and 5D black holes, and also for fermions.~\cite{OZ}

\section{ Dual Gravity Description of Noncommutative Super Yang-Mills}

If one introduces {\it constant background $B$ field} on the D-branes,
it produces {\it noncommutative SYM theory} on the branes,
and in an appropriate limit one can define noncommutative field theory
decoupled from gravity.~\cite{SW}
The generalization of AdS/CFT correspondence is possible.~\cite{MR,HI,CO1}
The theory can be described {\it either by noncommutative field theory
or dual gravity} which is asymptotically AdS but significantly deviates
from AdS in the short distance.
The gravity solution has nonextreme generalization with horizon, which
corresponds to field theory at finite temperature (the extreme case corresponds
to zero temperature).

Naively noncommutative space is ``discretized'' because of the ``uncertainty
relations'' arising from the noncommutativity $[x^\mu, x^\nu] = ic^{\mu\nu}$,
so that one would expect that {\it the degrees of freedom are less than the
ordinary theories.}
On the other hand, the interactions in noncommutative theories involve
higher derivatives (in the $\star$-product), so that one might expect
{\it opposite.} Which is correct?

By comparing thermodynamic quantities (energy, entropy and temperature)
computed in the gravity side for both commutative (AdS) and noncommutative
(non-AdS) cases, we can show that {\it the degrees of freedom in these
theories are the same in the leading order in the large $N$ limit} ($N$ is
the number of colors for $SU(N)$ theories).~\cite{CO1,CO2}
Thus the dual gravity description is quite useful to examine problems in
field theories.

\section{ Cardy-Verlinde Formula}

The holographic principle states that for a given volume $V$, the state of
maximal entropy is given by the largest black hole that fits inside $V$.
The microscopic entropy $S$ associated with the volume $V$ is less than the
Bekenstein-Hawking entropy:
\bea
S \leq \frac{\rm area}{4G}.
\ena
Verlinde observed that this bound is modified in the cosmological setting
and in arbitrary dimensions.~\cite{VE} His main observations are the
followings:
\begin{enumerate}
\item
Consider the space-time with the metric for the Einstein universe
\bea
ds^2 =-dt^2 +R^2d\Omega_n^2,
\label{eu}
\ena
where $d\Omega^2_n$ is the line element of a unit $n$-dimensional sphere.
The entropy of the CFT in this space-time can be reproduced in
terms of its total energy $E$ and Casimir energy $E_c$ by a generalized form
of the Cardy-Verlinde formula as
\bea
S=\frac{2\pi R}{n}\sqrt{E_c(2E-E_c)}.
\label{cvf}
\ena
(The original Cardy formula is for 2D CFT.)
\item
For an $(n+1)$-dimensional closed universe, the {FRW equations} are
\bea
&& H^2 =\frac{16 \pi G_n}{n(n-1)}\frac{E}{V} -\frac{1}{R^2}, \\
&& \dot{H}=-\frac{8 \pi G_n}{n-1}\left (\frac{E}{V} +p\right) +\frac{1}{R^2},
\ena
where $H=\dot{R}/R$ is the Hubble parameter,
dot stands for differentiation with respect to the proper time,
$E$ is the total energy of matter filling the universe,
$p$ the pressure,
$V= R^n \mbox{Vol}(S^n)$ the volume of the universe, and finally
$G_n$ is the $(n+1)$-dimensional gravitational constant.
The FRW equation can be related to three cosmological entropy bounds:
\begin{enumerate}
\item
{Bekenstein-Verlinde bound}:
\bea
S_{\rm BV}=\frac{2\pi}{n}ER.
\ena
(For a system with limited self-energy, the total entropy is less than
the product of energy and linear size of the system.)
\item
{Bekenstein-Hawking bound}:
\bea
S_{\rm BH}=(n-1)\frac{V}{4G_nR}.
\ena
(Black hole entropy is bounded by the area.)
\item
{Hubble  bound}:
\bea
S_{\rm H}=(n-1)\frac{HV}{4G_n}.
\ena
(Maximal entropy is produced by black holes of the size of Hubble horizon.)
\end{enumerate}
\end{enumerate}
At the critical point defined by $HR=1$, these three entropy bounds coincide
with each other.

Define $E_{\rm BH}$ such that
\bea
S_{\rm BH}=(n-1) V/4G_nR \equiv 2\pi E_{\rm BH} R/n.
\ena
The FRW equation then takes the form
\bea
S_{\rm H}=\frac{2\pi R}{n}\sqrt{E_{\rm BH}(2E-E_{\rm BH})},
\ena
which is of the same form as the Cardy-Verlinde formula~\p{cvf}!
Its maximum reproduces Hubble bound
\bea
S_H \leq \frac{2\pi R}{n} E.
\ena
Thus the FRW equation somehow knows the entropy of CFTs filling
the universe. This connection between the Cardy-Verlinde
formula and the FRW equation can be interpreted as a consequence of the
holographic principle.

Our purpose is to show that it is possible to extend this holographic
connection to the AdS Reissner-Nordstr\"om (RN) black hole background in
arbitrary dimensions.~\cite{CMO}

\subsection{ Bekenstein Bound in Arbitrary Dimensions}

The Bekenstein bound
\bea
S \le S_{\rm B}= 2\pi R E,
\label{bb}
\ena
is valid for a system with the limited self-gravity ({\it i.e.} if the
gravitational self-energy is negligibly small compared to its total energy).

It is known that the form of the Bekenstein bound~\p{bb} is independent
of the spatial dimensions, and that the $D (\ge 4)$-dimensional Schwarzschild
black hole satisfies the bound.
The bound is saturated even for a four-dimensional Schwarzschild black
hole which is a strongly self-gravitating object, but is no longer saturated
for $D>4$.

For charged objects with charge $Q$ in 4 dimensions, the Bekenstein
bound~\p{bb} is modified to
\bea
S \le S_{\rm B}=\pi (2ER -Q^2).
\ena
The question then arises: Does this form remain unchanged in arbitrary
dimensions ($D\ge 4$)?

Consider an $(n+2)$-dimensional Einstein-Maxwell theory with a
cosmological constant $\Lambda_{\pm} = \pm n(n+1)/2l^2$:
\newcommand{\g}{\bf G_n}
\bea
I = \frac{1}{16\pi \g}\int d^{n+2}x\sqrt{-g} \left ( {\cal R}
-F_{\mu\nu}F^{\mu\nu} - 2 \Lambda_{\pm}\right),
\ena
where ${\cal R}$is the curvature scalar,
$F$ the Maxwell field,
and $\g$ the gravitational constant in ($n+2)$ dimensions.
Let us discuss a spherically symmetric solution in this theory:
\bea
ds^2 = -f_\pm(r)dt^2 +f_\pm(r)^{-1}dr^2 +r^2 d\Omega_n^2,
\ena
where
\bea
&& F_{rt}=\frac{n\omega_n}{4}\frac{Q}{r^n}, \ \ \ \
 \omega_n=\frac{16\pi \g}{n \mbox{Vol}(S^n)},\\
&& f_{\pm}(r) = 1 -\frac{\omega_n M}{r^{n-1}} +\frac{n \omega^2_n Q^2}{8(n-1)
   r^{2(n-1)}} -\frac{2\Lambda_{\pm}r^2}{n(n+1)}.
\ena
where $\mbox{Vol}(S^n)$ is the volume of a unit $n$-sphere.

This solution is asymptotically de Sitter (dS) or anti-de Sitter (AdS)
depending on the cosmological constant $\Lambda_+$ or $\Lambda_-$.
We consider the de Sitter case in this subsection.

If we put $M=Q=0$, the solution describes dS space with a cosmological
horizon at $r=r_0\equiv \sqrt{l^2}$.
The cosmological horizon behaves like the black hole horizon, and it
has the thermodynamic entropy
\bea
S_0=\frac{r_0^n}{4\g}\mbox{Vol}(S^n).
\ena
In a general case with nonvanishing $M$ and $Q$, the solution describes
the geometry of a certain object with mass $M$ and electric charge $Q$
in dS space. The cosmological horizon will shrink due to the nonzero
$M$ and $Q$. This leads to the bound
\bea
S_m \le S_0 -S_c,
\ena
where $S_c$ is the cosmological horizon entropy when matter is present.

We now estimate $S_c$.
The cosmological horizon $r_c$ is given by the maximal root of the equation:
\bea
1-\frac{\omega_n M}{r_c^{n-1}} +\frac{n\omega_n^2 Q^2}{8(n-1)r_c^{2(n-1)}}
   -\frac{r_c^2}{r_0^2}=0.
\ena
This leads to
\bea
\frac{r_0^n}{r_c^n} =\left (1 -\frac{\omega_n M}{r_c^{n-1}}
  +\frac{n\omega_n^2 Q^2}{8 (n-1)r_c^{2(n-1)}} \right)^{-n/2}.
\ena
In the large cosmological horizon limit: $M/r_c^{n-1} \ll 1,Q^2/r_c^{2(n-1)}
\ll 1$, we get
\bea
S_m &\le & \frac{\mbox{Vol}(S^n)}{4\g} (r_0^n-r_c^n) \nonumber \\
 &\le& 2\pi r_c \left (M -\frac{2\pi {\g} Q^2}{(n-1)\mbox{Vol}(S^n) r_c^{n-1}}
  \right ).
\ena
The entropy reaches its maximum when the matter extends to the
cosmological horizon. If we replace $r_c$ by $R$ and $M$ by the proper energy
$E$, we obtain the entropy bound of the charged object in
arbitrary dimensions ($D=n+2 \ge 4 $):
\bea
S \le S_{\rm B}= 2\pi R \left (E -\frac{2\pi {\g} Q^2}{(n-1)
 \mbox{Vol}(S^n)R^{n-1}} \right).
\ena

Let us make various checks of this formula:
\begin{itemize}
\item
This bound is satisfied by RN black holes in arbitrary dimensions.
\item
When $Q^2=0$, this bound reproduces precisely the Bekenstein bound
for the neutral object in arbitrary dimensions.
\item
For $n=2$, the entropy bound reduces to the four-dimensional one.
\end{itemize}
These confirm the validity of the above bound.

\subsection{ AdS Reissner-Nordstr\"om Black Holes and Bekenstein-Verlinde
 bound}

In this subsection, we proceed to the AdS case of $\Lambda_-$.
Basically similar bound can be derived for charged case.

In this case, the cosmological horizon is absent, and
the solution describes the AdS RN black hole in arbitrary dimensions.
The black hole horizon $r_+$ is determined by the maximal root
of the equation $f_-(r_+)=0$.

In the spirit of the AdS/CFT correspondence, the thermodynamics of AdS RN
black holes corresponds to that for the boundary CFT.
We rescale the boundary metric of the solution so that it has the form of
Einstein universe~\p{eu}.

Thermodynamic quantities of the corresponding CFT in Einstein universe
are found to be
\bea
&& \mbox{energy}: E = \frac{l r_+^{n-1}}{R\omega_n}\left ( 1+\frac{r_+^2}{l^2}
       +\frac{n\omega_n^2 Q^2}{8(n-1)r_+^{2(n-1)}}\right), \\
&& \mbox{temperature}: T =\frac{l}{4\pi Rr_+}\left((n-1)
 +\frac{(n+1)r_+^2}{l^2}-\frac{n\omega_n^2 Q^2}{8r_+^{2(n-1)}} \right), \\
&& \mbox{chemical potential (charge $Q$)}: \Phi =\frac{nl\omega_n Q}{4(n-1)R
 r_+^{n-1}}, \\
&& \mbox{entropy}: S = \frac{r_+^n}{4\g}\mbox{Vol}(S^n), \\
&& \mbox{Gibbs free energy}: G = \frac{lr_+^{n-1}}{nR\omega_n}\left(
1-\frac{r_+^2}{l^2}
     -\frac{n\omega_n^2}{8(n-1)}\frac{Q^2}{r_+^{2(n-1)}}\right),
\ena
where $r_+$ is the horizon of the AdS RN black hole.

The entropy can be written in a form analogous to the Cardy-Verlinde formula
\bea
S =\frac{2\pi R}{n}\sqrt{E_c(2(E-E_q)-E_c}),
\ena
where
\bea
E_c = \frac{2lr_+^{n-1}}{\omega_n R}, \ \ \  E_q=\frac{1}{2}\Phi Q =
   \frac{l}{R}\frac{n\omega_n}{8(n-1)}\frac{Q^2}{r_+^{n-1}}.
\ena
Its maximum is
\bea
S_{\rm max} &=& \frac{2\pi R}{n}(E-E_q) \nonumber \\
            &=& \frac{2\pi}{n}\left (ER -\frac{nl\omega_n}{8(n-1)}
              \frac{Q^2}{r_+^{n-1}}\right),
\ena
at $E_c =E-E_q$.

A Bekenstein-Verlinde-like bound for a charged system is then obtained.
According to the AdS/CFT correspondence, the boundary space-time in which
the boundary CFT resides can be determined from the bulk metric, up to a
conformal factor. Rescale the boundary metric so that the radius $R$ becomes
the horizon radius $r_+$ of the black hole. The maximal entropy then gives
\bea
S_{\rm max}=\frac{2\pi R}{n}\left (E -\frac{nl\omega_n}{8(n-1)}\frac{Q^2}
    {R^n}\right).
\ena
or
\bea
S_{\rm BV} =\frac{2\pi R}{n}\left (E -\frac{2\pi {\g}l}{(n-1)}\frac{Q^2}{V}
        \right).
\ena

It may be slightly puzzling that bulk parameter $l$ appears in
the bound. In the AdS/CFT correspondence, the cosmological constant is
related to the 't Hooft coupling constant in the CFT. So there is no
contradiction with holographic principle!

\section{ Conclusions}

In this review we have tried to give rather intuitive picture of the
M-theory and other related recent developments in superstring theories,
starting with the introduction to string theories. The picture emerging from
this is that the string theory is never a theory of strings only,
but a theory of many extended objects which are intricately combined to
exhibit its appearance as various string theories on perturbative vacua.
In this view, D-branes play very significant roles and study of
their properties are expected to shed further light on the nature of
M-theory.

In addition, the D-brane physics is quite rich and interesting. They allow
nonperturbative study of gauge field theories, including noncommutative
theories (though this point was not discussed here).

It is extremely interesting and important to understand how M-theory unifies
all the string theories (or to be more precise, string vacua), and clarify
the dynamics implied by this theory. The problems include
\begin{itemize}
\item
how to formulate the M-theory itself precisely,
\item
how to understand the dynamics of compactification.
\end{itemize}
AdS/CFT correspondence (or Open/Closed string duality)
and noncommutative geometry might be important in this task.

\section*{Acknowledgements}

The author would like to thank the organizers of the international
workshop ``Braneworld - Dynamics of spacetime with boundary'' for giving
him the opportunity of summarizing the recent developments in string
theories and also presenting some of his results, as well as the
participants for stimulating discussions.
This work was supported in part by a Grant-in-Aid for Scientific
Research No. 12640270, and by a Grant-in-Aid on the Priority Area:
Supersymmetry and Unified Theory of Elementary Particles.

%\newpage

\end{document}